\documentstyle[preprint,aps]{revtex}
\input epsf.tex
\def\DESepsf(#1 width #2){\epsfxsize=#2 \epsfbox{#1}}
\begin{document}
\preprint{OITS-603}
\draft
\title{ On Measuring CP Violating Phase $\gamma$ Using Neutral $B$ Decays }
\author{N.G. Deshpande and Sechul Oh}
\address{Institute of Theoretical Science\\
 University of Oregon\\
 Eugene, OR 97403-5203, USA\\ 
 e-mail:desh@oregon.uoregon.edu\\
  sco@oregon.uoregon.edu}
\date{June, 1996}
\maketitle
\begin{abstract}

We propose two independent methods to determine the CKM phase $\gamma$ and the tree and penguin 
amplitudes using neutral $B$ decays, assuming that the phase $\beta$ is known.  
Based on flavor SU(3) symmetry and SU(3) breaking effect, 
one method uses the decay processes $B^0(t) \rightarrow \pi^0 K_S$,  
$B^0 \rightarrow \eta K^0 \; (K_S \rightarrow \pi \pi)$ and their charge-conjugate processes, 
and the other uses the processes $B^0(t) \rightarrow \pi^0 K_S$, $B^0 \rightarrow \pi^0 \pi^0$ and 
their charge-conjugate processes.  From SU(3) breaking consideration, the latter method would be 
more useful.  

\end{abstract}


\newpage

The Standard Model of three generations with the source for CP violation arising from the phases 
in the Cabibbo-Kobayashi-Maskawa (CKM) matrix is so far consistent with the experiment \cite{1}. 
An important way of verifying the CKM model is to measure the three angles 
$\alpha \equiv \mbox{Arg}[-(V_{td}V^*_{tb})/(V_{ub}^*V_{ud})]$,
$\beta \equiv \mbox{Arg}[-(V_{cd}V^*_{cb})/(V_{tb}^*V_{td})]$, and
$\gamma \equiv  \mbox{Arg}[-(V_{ud}V^*_{ub})/(V_{cb}^*V_{cd})]$ of the unitarity triangle of the 
CKM matrix independently in experiments and to check whether the sum of these three angles is 
equal to $180^o$, as it should be in the model.  
$B$ meson decays provide a fertile ground to carry out such a test \cite{2,3}. 
One class of methods to measure the CKM phases involve the measurements 
of CP asymmetries in time evolution of $B^0$ decays into CP eigenstates \cite{3}.
Since most decay processes get contributions from both tree and loop (penguin) effects, in order 
to measure the CKM phases without hadronic uncertainties, in many cases one needs additional 
information such as using relations based on isospin or flavor SU(3) symmetries 
\cite{4,5,6,7,8,9,10,11,12,13,14}.  
The phase $\beta$ can be determined unambiguously by measuring CP asymmetry in time evolution of 
$B \rightarrow \Psi K_S$ \cite{3}.  
The extraction of $\alpha$ involves the study of CP asymmetry in $B \rightarrow \pi \pi$ or
$B \rightarrow \rho \pi$ \cite{4,5}. 
If the penguin contributions are neglected, this extraction is straightforward.  However, if 
penguin diagrams make a significant contribution, then the interpretations of the results become 
complicated.   
The methods proposed in Refs. \cite{4,5} are valid even if the strong penguin contributions are 
included, and the inclusion of the electroweak penguin contributions makes only a small 
error in those $\alpha$ determinations \cite{10,13}.  
A few other methods using $B \rightarrow \pi \pi$ decays to determine $\alpha$ have been 
suggested as well \cite{14}.  

For the phase $\gamma$ determination, several methods using $|\Delta S| =$ 1 nonleptonic $B$ decays 
have been proposed \cite{6,7,8}.  Some assumed that the electroweak penguin contributions could be 
neglected.  
It was shown that this assumption is badly violated for a top quark mass of order 170 GeV \cite{10}.  
In $|\Delta S| =$ 1 $B$ decays the penguin contributions are 
enhanced by a factor of $|V_{tb} V_{ts}^*| /|V_{ub} V_{us}^*| \approx$ 50 compared to the tree 
contributions so that the strong penguins dominate and the electroweak penguin contributions are 
comparable to the tree ones, while in $\Delta S =$0 decays the penguin effects are much smaller 
than the leading tree contributions and so the effects from the electroweak penguins can be 
safely neglected.  
To overcome the difficulties associated with the electroweak contamination, some newly proposed 
methods have used the certain relations between decay amplitudes for nonleptonic 
$B$ decays including $|\Delta S| =$ 1 decays, based on flavor SU(3) symmetry \cite{11}.  
In a recent paper \cite{15} Gronau and Rosner suggested an interesting method to determine 
simultaneously 
both $\alpha$ and $\gamma$, using the decays $B^0(t) \rightarrow \pi^+ \pi^-$, 
$B^0 \rightarrow \pi^- K^+$, $B^+ \rightarrow \pi^+ K^0$ $(K_S \rightarrow \pi^+ \pi^-)$ and 
their charge-conjugate decays, and  employing flavor SU(3) symmetry.  

In this letter we present two independent methods to determine the phase $\gamma$ using the 
time-dependent rate measurement of $B^0(\bar B^0) \rightarrow \pi^0 K_S$ and the time-integrated 
rate measurement of $B^0 \rightarrow \eta K^0(\bar B^0 \rightarrow \eta \bar K^0)$ 
(where $K^0(\bar K^0) \rightarrow K_S \rightarrow \pi \pi$) or 
$B^0(\bar B^0) \rightarrow \pi^0 \pi^0$, assuming that the phase $\beta$ is known.  
The phase $\beta$ is expected to be measured very cleanly through the process 
$B \rightarrow \Psi K_S$.    
Our methods are based on flavor SU(3) symmetry and considering SU(3) breaking effects.  
These methods are free from the electroweak penguin contamination problem and can determine 
the tree and penguin amplitudes of the involved decays as well.  

The effective Hamiltonian up to one loop level in electroweak interaction for hadronic $B$ decays 
can be written as 
\begin{eqnarray}
 H_{\Delta B =1} = {4 G_{F} \over \sqrt{2}} \left[V_{ub}V^{*}_{uq} (c_1 O^{u}_1 +c_2 O^{u}_2) 
   + V_{cb}V^{*}_{cq} (c_1 O^{c}_1 +c_2 O^{c}_2)
   - V_{tb}V^{*}_{tq} \sum_{i=3}^{12} c_{i} O_{i} \right] + H.C. ,
\label{HAM}
\end{eqnarray} 
where $O_{i}$'s are defined as 
\begin{eqnarray}
 &O&^{f}_{1} = \bar q_{\alpha} \gamma_{\mu} L f_{\beta} \bar f_{\beta} \gamma^{\mu} L b_{\alpha} ,
  \ \  O^{f}_{2} = \bar q \gamma_{\mu} L f \bar f \gamma^{\mu} L b ,           \nonumber \\ 
 &O&_{3(5)} = \bar q \gamma_{\mu} L b \Sigma \bar q^{\prime} \gamma^{\mu} L(R) q^{\prime} ,   
  \ \ O_{4(6)} = \bar q_{\alpha} \gamma_{\mu} L b_{\beta} \Sigma \bar q^{\prime}_{\beta} 
       \gamma^{\mu} L(R) q^{\prime}_{\alpha}  ,                         \nonumber \\ 
 &O&_{7(9)} = {3 \over 2} \bar q \gamma_{\mu} L b \Sigma e_{q^{\prime}} \bar q^{\prime} 
       \gamma^{\mu} R(L) q^{\prime} ,        
  \ \ O_{8(10)} ={3 \over 2} \bar q_{\alpha} \gamma_{\mu} L b_{\beta} \Sigma e_{q^{\prime}} 
       \bar q^{\prime}_{\beta} \gamma^{\mu} R(L) q^{\prime}_{\alpha} ,      \nonumber \\ 
&O&_{11} ={g_{s}\over{32\pi^2}}m_{b}\bar q \sigma_{\mu \nu} RT_{a}b 
G_{a}^{\mu \nu} \;,\;\;
O_{12} = {e\over{32\pi^2}} m_{b}\bar q \sigma_{\mu \nu}R b 
F^{\mu \nu} \;,
\end{eqnarray} 
where $L(R) = (1 \mp \gamma_5)/2$, $f$ can be $u$ or $c$ quark, $q$ can be $d$ or $s$ quark, 
and $q^{\prime}$ is summed over $u$, $d$, $s$, and $c$ quarks.  $\alpha$ and $\beta$ are 
the color indices.  
$T^{a}$ is the SU(3) generator with the normalization $Tr(T^{a} T^{b}) = \delta^{ab}/2$.  
$G^{\mu \nu}_{a}$ and $F_{\mu \nu}$ are the gluon and photon field strength, respectively. $c_i$ 
are the Wilson Coefficients (WC).  $O_1$, $O_2$ are the tree level and QCD corrected operators.  
$O_{3-6}$ are the gluon induced strong penguin operators.  $O_{7-10}$ are the electroweak penguin 
operators due to $\gamma$ and $Z$ exchange, and ``box'' diagrams at loop level. The operators 
$O_{11,12}$ are the dipole penguin operators. 
The WC's $c_{i}$ at a particular scale $\mu$ are obtained by first calculating the WC's at $m_{W}$ 
scale and then using the renormalization group equation to evolve them to $\mu$.  
In the above we have neglected small contributions from $u$ and $c$ quark loops 
proportional to $V_{ub} V_{uq}^*$.  

We can always parameterize the decay amplitude of $B$ that arises from quark subprocess 
$b \rightarrow u \bar u q$ as 
\begin{eqnarray}
\bar A = <final\;state|H_{eff}^q|B> = V_{ub}V^*_{uq} T(q) + V_{tb}V^*_{tq}P(q)\;,
\end{eqnarray}
where $T(q)$ contains $tree$ contributions, while $P(q)$ contains purely 
$penguin$ contributions. 

SU(3) relations for $B$ decays have been studied by several authors \cite{11,16,17}.
The operators $O_{1,2}$, $O_{3-6, 11,12}$, and $O_{7-10}$ transform under SU(3)
symmetry as $\bar 3_a + \bar 3_b +6 + \overline {15}$,
$\bar 3$, and $\bar 3_a + \bar 3_b +6 + \overline {15}$, respectively. In general, we can
write the SU(3) invariant amplitude for $B$ decay to two octet pseudoscalar mesons. 
For the $T$ amplitude, for example, we have 
\begin{eqnarray}
T&=& A_{(\bar 3)}^TB_i H(\bar 3)^i (M_l^k M_k^l) + C^T_{(\bar 3)}
B_i M^i_kM^k_jH(\bar 3)^j \nonumber\\
&+& A^T_{(6)}B_i H(6)^{ij}_k M^l_jM^k_l + C^T_{(6)}B_iM^i_jH(6
)^{jk}_lM^l_k\nonumber\\
&+&A^T_{(\overline {15})}B_i H(\overline {15})^{ij}_k M^l_jM^k_l +
C^T_{(\overline
{15})}B_iM^i_j
H(\overline {15} )^{jk}_lM^l_k\;,
\end{eqnarray}
where $B_i = (B^-, \bar B^0, \bar B^0_s)$ is a SU(3) triplet, $M_{i}^j$ is the SU(3) pseudoscalar 
octet, and the matrices $H$ represent the transformation properties of the operators $O_{1-12}$.
$H(6)$ is a traceless tensor that is antisymmetric on its upper indices, and
$H(\overline {15} )$ is also a traceless tensor but is symmetric on its upper indices. 
We can easily see that the strong and dipole penguin operators only contribute to 
$A_{(\bar 3)}$ and $C_{(\bar 3)}$. 

For $q=d$, the non-zero entries of the $H$ matrices are given by
\begin{eqnarray}
H(\bar 3)^2 &=& 1\;,\;\;
H(6)^{12}_1 = H(6)^{23}_3 = 1\;,\;\;H(6)^{21}_1 = H(6)^{32}_3 =
-1\;,\nonumber\\
H(\overline {15} )^{12}_1 &=& H(\overline {15} )^{21}_1 = 3\;,\; H(\overline
{15} )^{22}_2 =
-2\;,\;
H(\overline {15} )^{32}_3 = H(\overline {15} )^{23}_3 = -1\;.
\end{eqnarray}
For $q = s$, the non-zero entries are
\begin{eqnarray}
H(\bar 3)^3 &=& 1\;,\;\;
H(6)^{13}_1 = H(6)^{32}_2 = 1\;,\;\;H(6)^{31}_1 = H(6)^{23}_2 =
-1\;,\nonumber\\
H(\overline {15} )^{13}_1 &=& H(\overline {15} ) ^{31}_1 = 3\;,\; H(\overline
{15} )^{33}_3 =
-2\;,\;
H(\overline {15} )^{32}_2 = H(\overline {15} )^{23}_2 = -1\;.
\end{eqnarray}
In terms of the SU(3) invariant amplitudes, the decay amplitudes $T(\pi^0 \bar K^0)$, 
$T(\eta_8 \bar K^0)$ and $T(\pi^0 \pi^0)$ for $\bar B^0 \rightarrow \pi^0 \bar K^0$, 
$\eta_8 \bar K^0$, and $\pi^0 \pi^0$ are given by
\begin{eqnarray}
 T(\pi^0 \bar K^0) &=& - {1 \over \sqrt{2}} ( C^T_{(\bar 3)}
  - A^T_{(6)} + C^T_{(6)} - A^T_{(\overline {15} )} - 5 C^T_{(\overline {15})})\;,\nonumber\\
 T(\eta_8 \bar K^0) &=& - {1 \over \sqrt{6}} ( C^T_{(\bar 3)}
  - A^T_{(6)} + C^T_{(6)} - A^T_{(\overline {15} )} - 5 C^T_{(\overline {15})})\;,\nonumber\\
 T(\pi^0 \pi^0) &=& {1 \over \sqrt{2}} ( 2A^T_{(\bar 3)} + C^T_{(\bar 3)}
  - A^T_{(6)} + C^T_{(6)} + A^T_{(\overline {15} )} - 5 C^T_{(\overline {15})}).
\label{TTT}
\end{eqnarray}
We also have similar relations for the amplitude $P(q)$. The corresponding
SU(3) invariant amplitudes will be denoted by $A^P_i$ and $C^P_i$.

If we neglect small contribution from the annihilation terms $A^T_i$ and $A^P_i$, we obtain in the 
SU(3) limit the following relations: 
\begin{eqnarray} 
 \sqrt{2} T(\pi^0 \bar K^0) &=& \sqrt{6} T(\eta_8 \bar K^0) = -\sqrt{2} T(\pi^0 \pi^0),\nonumber \\
 \sqrt{2} P(\pi^0 \bar K^0) &=& \sqrt{6} P(\eta_8 \bar K^0) = -\sqrt{2} P(\pi^0 \pi^0),
\label{TP}
\end{eqnarray} 
which imply in the SU(3) limit the relation: 
\begin{eqnarray} 
 \sqrt{2} \bar A(\bar B^0 \rightarrow \pi^0 \bar K^0) = \sqrt{6} \bar A(\bar B^0 \rightarrow 
   \eta_8 \bar K^0).
\end{eqnarray}
From relations (\ref{TP}), in the exact SU(3) limit the decay amplitudes for 
$\bar B^0 \rightarrow \pi^0 \bar K^0$, $\eta_8 \bar K^0$ and $\pi^0 \pi^0$ can be 
parameterized as 
\begin{eqnarray} 
 \bar A_{\pi K} &\equiv& <\pi^0 \bar K^0|H|\bar B^0> = 
  {1\over \sqrt{2}} (T e^{-i\gamma} e^{i\delta_T} + P e^{i\delta_P}),  \nonumber \\ 
 \bar A_{\eta K} &\equiv& <\eta_8 \bar K^0|H|\bar B^0> = 
  {1\over \sqrt{6}} (T e^{-i\gamma} e^{i\delta_T} + P e^{i\delta_P}),  \nonumber \\ 
 \bar A_{\pi \pi} &\equiv& <\pi^0 \pi^0|H|\bar B^0> = 
  - {1\over \sqrt{2}} (v_u T e^{-i\gamma} e^{i\delta_T} + v_t P e^{i\beta} e^{i\delta_P}),  
\label{AAA}
\end{eqnarray} 
where $v_u \equiv |V_{ud}| / |V_{us}|$ and $v_t \equiv |V_{td}| / |V_{ts}|$.  

Now we consider SU(3) breaking effects, assuming that the strong phases $\delta_T$ and $\delta_P$ 
are unaffected by SU(3) breaking.  Eqs.(\ref{AAA}) can be rewritten as 
\begin{eqnarray} 
 \sqrt{2} \bar A_{\pi K} &=& T e^{-i\gamma} e^{i\delta_T} + P e^{i\delta_P},  
\label{pK} \\
 \sqrt{6} \bar A_{\eta K} &=& T^{\prime} e^{-i\gamma} e^{i\delta_T} + P^{\prime} e^{i\delta_P}, 
\label{eK} \\
 \sqrt{2} \bar A_{\pi \pi} &=& - v_u T^{\prime \prime} e^{-i\gamma} e^{i\delta_T} 
     - v_t P^{\prime \prime} e^{i\beta} e^{i\delta_P}.   
\label{pp} 
\end{eqnarray} 
While penguin amplitudes are allowed to have full breaking effects, we will use factorization 
for tree amplitudes to estimate SU(3) breaking.  For the moment we will ignore the breaking effect 
due to $\eta$-$\eta^{\prime}$ mixing.  
Using the effective Hamiltonian (\ref{HAM}) for calculation we find
\begin{eqnarray} 
 a &\equiv& {T^{\prime} \over T} 
   ={f_{\eta_8} [f^+_{BK}(m^2_{\eta_8}) (m^2_B -m^2_K) + f^-_{BK}(m^2_{\eta_8}) m^2_{\eta_8}] \over 
      f_{\pi} [f^+_{BK}(m^2_{\pi}) (m^2_B -m^2_K) + f^-_{BK}(m^2_{\pi}) m^2_{\pi}] }
   = 1.34, \;\nonumber\\ 
 b &\equiv& {T^{\prime \prime} \over T} 
   ={ f^+_{B\pi}(m^2_{\pi}) (m^2_B -m^2_{\pi}) + f^-_{B\pi}(m^2_{\pi}) m^2_{\pi}  \over 
      f^+_{BK}(m^2_{\pi}) (m^2_B -m^2_K) + f^-_{BK}(m^2_{\pi}) m^2_{\pi} } 
   = 0.89 ,   
\label{ab}
\end{eqnarray} 
For numerical values we have used $m_{\eta_8}=$ 613 MeV, the decay constants 
$f_{\pi}=$ 133 MeV and $f_{\eta_8}=$ 176 MeV, and the form factors $f^{\pm}_{B\pi}$ and 
$f^{\pm}_{BK}$ calculated in Refs. \cite{18}.  
We see that the SU(3) breaking for tree amplitudes is larger in the former case because of 
large breaking effect for $f_{\eta_8} / f_{\pi}$.   
 
First we will discuss a method to determine the parameters $\gamma$, $\delta$, $T$, $P$, and 
$P^{\prime}$ in Eqs.(\ref{pK}) and (\ref{eK}) assuming that $\beta$ is known. 
From the Eqs.(\ref{pK}) and (\ref{eK}), the following relations can be obtained: 
\begin{eqnarray} 
 \tilde X &\equiv& |A_{\pi K}|^2 + |\bar A_{\pi K}|^2 
   = T^2 +P^2 - 2 T P \cos{\delta} \cos{\gamma} , \nonumber \\
 \tilde Y &\equiv& |A_{\pi K}|^2 - |\bar A_{\pi K}|^2 
   = 2 TP \sin{\delta} \sin{\gamma} , \;\nonumber \\
 \tilde Z &\equiv& Im \left( e^{2i\beta} A_{\pi K} \bar A^{*}_{\pi K} \right) 
   = \left[ T^2 \sin{2(\beta + \gamma)} + P^2 \sin(2\beta) 
           + 2 TP \cos{\delta} \sin{(2\beta + \gamma)} \right]/2 
  \nonumber \\
 \tilde U &\equiv& 3 \left( |A_{\eta K}|^2 + |\bar A_{\eta K}|^2 \right) 
   = a^2 T^2 +{P^{\prime}}^2 - 2 a T P^{\prime} \cos{\delta} \cos{\gamma} , \nonumber \\
 \tilde V &\equiv& 3 \left( |A_{\eta K}|^2 - |\bar A_{\eta K}|^2 \right) 
   = 2 a T P^{\prime} \sin{\delta} \sin{\gamma}, 
\label{XYZ}
\end{eqnarray} 
where $\delta$ is defined by $\delta = \delta_T - \delta_P$.  
$A_{ij}$'s denote the CP-conjugate amplitudes of $\bar A_{ij}$'s.  
The quantities $\tilde U$ and $\tilde V$ can be determined by measurement of the 
time-integrated decay rates for $B^0 \rightarrow \eta_8 K^0$ and 
$\bar B^0 \rightarrow \eta_8 \bar K^0$, where $K^0(\bar K^0)$ can be observed by 
$K_S \rightarrow \pi \pi$. 
To determine the 
quantities $\tilde X$, $\tilde Y$, and $\tilde Z$, one can measure the time-dependent rates for 
initially pure $B^0$ or 
$\bar B^0$ states to decay into $\pi^0 K_S$ at time $t$, which are given by \cite{7}
\begin{eqnarray} 
 \Gamma &(& B^0(t) \rightarrow \pi^0 K_S )    \nonumber \\ 
   &=& e^{-\Gamma t} \left[|A_{\pi K}|^2 \cos^2 \left({\Delta m \over 2}t \right) 
   + |\bar A_{\pi K}|^2 \sin^2 \left( {\Delta m \over 2}t \right)   
   + Im \left( e^{2i\beta} A_{\pi K} \bar A^*_{\pi K} \right) 
        \sin \left(\Delta m t \right) \right],  \nonumber \\ 
 \Gamma &(& \bar B^0(t) \rightarrow \pi^0 K_S )   \nonumber \\ 
   &=& e^{-\Gamma t} \left[ |A_{\pi K}|^2 \sin^2 \left( {\Delta m \over 2}t \right) 
   + |\bar A_{\pi K}|^2 \cos^2 \left( {\Delta m \over 2}t \right)    
   - Im \left( e^{2i\beta} A_{\pi K} \bar A^*_{\pi K} \right) 
        \sin \left(\Delta m t \right) \right], 
\end{eqnarray} 
where $-\infty \leq t \leq +\infty$.  
Although the branching ratio for $B \rightarrow \pi K$ is expected to be small, it will be 
partially compensated by good detection efficiency for $K_S$ (like the case of 
$B \rightarrow \psi K_S$) as referred in Ref.\cite{7}.  
Measurement of these time-dependent decay rates gives sufficient information to determine 
$|A_{\pi K}|$, $|\bar A_{\pi K}|$, and $Im( e^{2i\beta} A_{\pi K} \bar A^{*}_{\pi K} )$.  

Now with the known five quantities $\tilde X$, $\tilde Y$, $\tilde Z$, $\tilde U$, and $\tilde V$, 
it is straightforward to determine all five parameters $\gamma$, $\delta$, $T$, $P$, and 
$P^{\prime}$ in Eqs.(\ref{XYZ}) 
up to discrete ambiguities, if we assume that $\beta$ is known.  
It is easy to show that the following relations hold:
\begin{eqnarray} 
 T^2 &=& \tilde F P^2 + \tilde G,   \nonumber \\
 P^{\prime} &=& \left({\tilde V \over a \tilde Y} \right) P,  \nonumber \\ 
 \sin{\delta} &=&  \tilde Y / (2 TP \sin{\gamma}),  
\end{eqnarray} 
where $\tilde F = (\tilde V/\tilde Y) (1 - \tilde V/(a^2 \tilde Y) ) / 
(a^2 - \tilde V/ \tilde Y)$ and 
$\tilde G = (\tilde U - \tilde V \tilde X/ \tilde Y) / (a^2 - \tilde V/ \tilde Y)$. 

Using Eq.(\ref{pp}) instead of Eq.(\ref{eK}) and following similar procedure to the one shown 
above, we can have another independent method to determine the phase $\gamma$, 
assuming that $\beta$ is known. 
In this case, since $v_t$ is largely unknown, the parameter $v_t P^{\prime\prime}$ can 
be determined instead of $P^{\prime\prime}$ itself.  
Measurement of the rates of processes $B^0 \rightarrow \pi^0 \pi^0$ and 
$\bar B^0 \rightarrow \pi^0 \pi^0$ is needed to determine the following quantities $\tilde R$ and 
$\tilde S$:
\begin{eqnarray} 
 \tilde R &\equiv& |A_{\pi\pi}|^2 + |\bar A_{\pi\pi}|^2 
   = v_u^2 b^2 T^2 +(v_t P^{\prime\prime})^2 
    +2 v_u bT (v_t P^{\prime\prime}) \cos{\delta} \cos{(\beta+\gamma)}, \nonumber \\
 \tilde S &\equiv& |A_{\pi\pi}|^2 - |\bar A_{\pi\pi}|^2 
   = -2 v_u bT (v_t P^{\prime\prime}) \sin{\delta} \sin{(\beta+\gamma)}, 
\label{RS}
\end{eqnarray} 
Combining Eqs.(\ref{RS}) with the first three equations of Eqs.(\ref{XYZ}) including the quantities 
$\tilde X$, $\tilde Y$, and $\tilde Z$ known from measurement of the time-dependent $B^0$ 
and $\bar B^0$ decay rates to $\pi^0 K_S$,  one can determine all the unknown parameters $\gamma$, 
$\delta$, $T$, $P$, and $v_t P^{\prime\prime}$.  
We remark in passing that if we use the CKM phase $\alpha = \pi -\beta -\gamma$, $v_t$ can be 
determined by the relation: 
\begin{eqnarray} 
 v_u v_t =  \sin{\gamma} / \sin{\alpha}.
\end{eqnarray} 

In the first method shown above, if we consider the $\eta$-$\eta^{\prime}$ mixing effect, 
$A_{\eta K} \equiv A(B^0 \rightarrow \eta_8 K^0)$ can be determined by the relation: 
\begin{eqnarray} 
 A(B^0 \rightarrow \eta_8 K^0) = A(B^0 \rightarrow \eta K^0) \cos{\theta} 
                                + A(B^0 \rightarrow \eta^{\prime} K^0) \sin{\theta},  
\end{eqnarray} 
where $\theta \approx$ $20^o$ \cite{19} is the $\eta$-$\eta^{\prime}$ mixing angle.  
The decay amplitudes $A(B^0 \rightarrow \eta K^0)$ and $A(B^0 \rightarrow \eta^{\prime} K^0)$ 
can be obtained from experiments.  In principle, we need to know the relative phase of these 
two amplitudes.  However since $\sin{\theta}$ is small, this phase is crucial only if 
$A(B^0 \rightarrow \eta^{\prime} K^0)$ is much larger than $A(B^0 \rightarrow \eta K^0)$.  
As we see in Eqs.(\ref{ab}), the SU(3) breaking effect for tree amplitude is also much smaller  
(about 11$\%$) in the second method using the decay $B^0 \rightarrow \pi^0 \pi^0$ 
instead of the decay $B^0 \rightarrow \eta_8 K^0$.  So one would expect that the second method 
is more useful.  
Even though we have used factorization approximation to determine SU(3) breaking effects, these 
calculations should be much better for the tree amplitudes as argued in Ref.\cite{15}.  
Although the tree contributions here are color suppressed, only the ratio of two tree 
amplitudes is involved.  Thus, the coefficients of the operators that are sensitive to color 
factor $N_C$ do not have to be known precisely for a reliable estimate of the ratio.  

In summary, we have proposed two independent methods to determine the phase $\gamma$ and 
the tree and penguin amplitudes of the involved neutral $B$ decays, 
assuming that the phase $\beta$ is known by the future experiment, 
for instance, using the decay process $B \rightarrow \Psi K_S$. 
These methods are free from electroweak penguin contamination problem.    
From SU(3) breaking consideration, the second method shown would be more useful.  

This work was supported in part by the Department of Energy Grant No. DE-FG06-85ER40224. 
We would like to thank Xiao-Gang He for helpful discussions.


\begin{references}
\bibitem{1} N. Cabibbo, Phys. Rev. Lett. {\bf 10}, 531 (1963); 
 M. Kobayashi and T. Maskawa, Prog. Theor. Phys. {\bf 49}, 652 (1973).

\bibitem{2} I.I. Bigi and A.I.Sanda, Nucl. Phys. {\bf B193}, 85 (1981); {\bf 281}, 41 (1987); 

\bibitem{3}   For a review see, Y. Nir and H.R. Quinn, in $B \ \ Decays$, edited by S. Stone, p.520 
 (World Scientific, Singapore, 2nd ed., 1994); I. Dunietz, $ibid$, p.550.
 A.J. Buras, Nuclear Instr. and Methods {\bf A368}, 1 (1995); M. Gronau, $ibid$, 21.  

\bibitem{4} M. Gronau and D. London, Phys. Rev. Lett. {\bf 65}, 3381 (1990).

\bibitem{5} A.E. Snyder and H.R. Quinn, Phys. Rev. {\bf D48}, 2139 (1993). 

\bibitem{6} H.J. Lipkin, Y. Nir, H.R. Quinn, and A.E. Snyder, Phys. Rev. {\bf D44}, 1454 (1991). 

\bibitem{7} Y. Nir and H.R. Quinn, Phys. Rev. Lett. {\bf 67}, 541 (1991). 

\bibitem{8} M. Gronau, D. London and J. Rosner, Phys. Rev. Lett. {\bf 73}, 21(1994). 

\bibitem{9} M. Gronau, O. Hernandez, D. London and J. Rosner, Phys. Rev. {\bf D50}, 4529(1994); 
 Phys. Lett. {\bf B333}, 500(1994);

\bibitem{10} N.G. Deshpande and Xiao-Gang He, Phys. Rev. Lett. {\bf 74}, 26 (1995); $ibid$ Erratum, 
 4099(1995).

\bibitem{11} M. Gronau, O. Hernandez, D. London and J. Rosner, Preprint, EFI-95-15, hep-ph/9504327, 
 1995;  
 N.G. Deshpande and Xiao-Gang He, Phys. Rev. Lett. {\bf 75}, 3064 (1995)

\bibitem{12} A.S. Dighe, Preprint, EFI-95-52, hep-ph/9509287, 1995; 
 M. Gronau and J. Rosner, Phys. Rev. {\bf D53}, 2516 (1996); 
 A.S. Dighe, M. Gronau and J. Rosner, Phys. Lett. {\bf B367}, 357 (1996).
 A.J. Buras and R. Fleischer, Phys. Lett. {\bf B365}, 390 (1996); 
 R. Fleischer, Phys. Lett. {\bf B365}, 399 (1996).

\bibitem{13} N.G. Deshpande, Xiao-Gang He and Sechul Oh, Preprint, OITS-593, hep-ph/9511462, 1995 
 (submitted to Z. Phys. C).  

\bibitem{14} A.J. Buras and R. Fleischer, Phys. Lett. {\bf B360}, 138(1995); 
 G. Kramer, W.F. Palmer and Y.L. Wu, Preprint, DESY 95-246, hep-ph/9512341, 1995; 
 N.G. Deshpande, Xiao-Gang He and Sechul Oh, Preprint, OITS-598, hep-ph/9604336, 1996 
 (to be published in Phys. Lett. B) 

\bibitem{15} M. Gronau and J. Rosner, Phys. Rev. Lett. {\bf 76}, 1200(1996).  

\bibitem{16} D. Zeppenfeld, Z. Phys. {\bf C8}, 77(1981); M. Savage and M. Wise, Phys. Rev. 
 {\bf D39}, 3346(1989); $ibid$ {\bf D40}, Erratum, 3127(1989).

\bibitem{17} N.G. Deshpande and Xiao-Gang He, Phys. Rev. Lett. {\bf 75}, 1703 (1995). 

\bibitem{18}  M. Bauer, B. Stech and M. Wirbel, Z. Phys. {\bf C34}, 103 (1987); 
 A. Deandrea, N. Di Bartolomeo, R. Gatto, and G. Nardulli, Phys. Lett. {\bf B318}, 549 (1993).  

\bibitem{19} Particle Data Group, Phys. Rev. {\bf D50}, 1173 (1994).  

\end{references}
\end{document}